\documentclass[%
reprint,
 amsmath,amssymb,
 aps,
]{revtex4-1}
\usepackage{graphicx}
\usepackage{braket}
\usepackage{bm}
\usepackage{amsmath}
\begin{document}
\title{Characteristic $\alpha$ and $^6\mathrm{He}$ decays of the linear-chains in $^{16}$C}
\author{T. Baba$^1$ and M. Kimura$^{1,2}$}
\affiliation{$^1$Department of Physics, Hokkaido University, 060-0810 Sapporo, Japan\\
$^2$Reaction Nuclear Data Centre, Faculty of Science, Hokkaido University, 060-0810 Sapporo, Japan}
\date{\today}

\begin{abstract}
 The linear-chain states of $^{16}$C and their decay modes are theoretically
 investigated by using the antisymmetrized molecular dynamics. It is found that the
 positive-parity  linear-chain states have the $(3/2^-_\pi)^2(1/2^-_\sigma)^2$
 configuration and primary decay to the $^{12}$Be($2^+_1$) as well as
 to the $^{12}$Be(g.s.) by the $\alpha$ particle emission.
 Moreover, we show that they also decay to the $^6{\rm He}+{}^{10}{\rm Be}$ channel.
 In the negative-parity states, it is found that two types of
 the linear-chains exist. One has the valence neutrons occupying the molecular-orbits
 $(3/2^-_\pi)^2(1/2^-_\sigma)(3/2^+_\pi)$, while the other's configuration cannot be explained
 in terms of the molecular orbits because of the strong parity mixing.
 Both configurations constitute the rotational bands with large moment of inertia and
 intra-bands $E2$ transitions.
 Their $\alpha$ and ${}^{6}{\rm He}$ reduced widths are sufficiently large to be
 distinguished from other non-cluster states although they are smaller than those of the
 positive-parity linear-chain.
\end{abstract}

\maketitle
\section{introduction}
%Some kinds of the $\alpha$-cluster configuration are well known in light nuclei.
%For example, the Hoyle state, the $0^+_2$ state of $^{12}$C, is the most famous clustering configuration, which is considered as the dilute gas-like cluster.
%One of the most exotic cluster is the linear-chain configuration in which 3$\alpha$ particles are linearly aligned.
%The linear-chain configuration has been searched in the excited states of $^{12}$C by many works since it was suggested in 1950's \cite{mori56}. 
%At present, it is considered that the $0^+_3$ state of $^{12}$C is the bent-armed linear-chain which is unstable against the bending motion \cite{enyo97,neff07}.

%In these decades neutron-rich C isotopes have attracted much interest as new candidates of the linear-chain.
%It is expected that the valence neutrons play a glue-like role and stabilize the linear-chain against the bending motion \cite{itag01}.
%It was pointed out that the motion of the valence neutron can be qualitatively interpreted in terms of the molecular orbits analogous to the Be isotopes \cite{seya81,oert96,yeny99,itag00,itag02,amd1,amd2}, which are called $\pi$- and $\sigma$-orbits.
%Naturally, this mechanism is extent to the linear-chain configurations in C isotopes.
%Indeed, some theoretical studies predict the rotational bands of C isotopes \cite{oert03,oert04,itag06,suha10,maru10,furu11,suha11,baba14,zhao15,kimu16,baba16}.

Recent years have seen many important experimental and theoretical studies for the linear-chain
states (linearly aligned 3$\alpha$ particles) in $^{14}$C
\cite{soic03,mili04,pric07,haig08,suha10,suha11,free14,ebran14,frit16,tian16,baba16,yoshi16,yama17,li17,baba17,ebran17}
and $^{16}$C
\cite{itag01,gree02,bohl03,ashw04,maru10,baba14,dell16}. 
In these C isotopes several theoretical studies predicted the existence of linear-chain states with the valence neutrons playing a glue-like role to stabilize the extreme shape.
The antisymmetrized molecular dynamics (AMD) calculations for $^{14}$C \cite{suha10,baba16,baba17} predicted a positive-parity rotational band with the linear-chain configuration having the $\pi$-bond valence neutrons.
The calculations also suggested a unique decay pattern of the $\pi$-bond linear-chain, {\it i.e.}, it decays not only to the ${}^{10}{\rm Be}(0^+_1)$ but also to the ${}^{10}{\rm Be}(2^+_1)$ by the $\alpha$ particle emission.
It was found that the energies, moment of inertia and decay pattern of the resonances observed by the $\alpha+{}^{10}{\rm Be}$ elastic scattering \cite{free14,frit16,yama17} reasonably agree with the predicted $\pi$-bond linear-chain.
Therefore, the $\pi$-bond linear-chain formation in ${}^{14}{\rm C}$ looks confidential.
In addition to the $\pi$-bond linear-chain, we also predicted another linear-chain which has $\sigma$-bond neutrons.
This band should have different decay pattern as it will dominantly decay to the ${}^{10}{\rm Be}(0^+_2)$ and ${}^{10}{\rm Be}(2^+_3)$.
Although the experimental information is not enough, the resonances observed by the $^{9}{\rm Be}(^{9}{\rm Be}, \alpha+{}^{10}{\rm Be})\alpha$ reaction \cite{li17} looks promising candidate for the $\sigma$-bond linear-chain.

The advances in the study of $^{14}$C naturally motivate us to study the linear-chains in neutron-rich C isotopes.
In particular, we expect the linear-chain states should also exist in $^{16}$C \cite{baba14}, because both of $\pi$- and $\sigma$-bonding orbits are simultaneously occupied by valence neutrons.
Actually, the molecular-orbital model calculation \cite{itag01} predicted that $^{16}$C has the most substantial linear-chain among the C isotopes.
And, our previous work \cite{baba14} predicted a positive-parity linear-chain band built on the $0^+$ state at 15.5 MeV, which should be verified experimentally.
In this work, for further experimental study, we provide additional  theoretical informations.
The first is negative-parity states.
In the case of $^{14}$C, the experiments reported the negative-parity resonances \cite{free14,frit16,yama17}.
Therefore, theoretical predictions will be needed for the negative-parity resonances in  $^{16}$C.
The second is decay mode of the linear-chain configuration.
Because the linear-chain band of $^{16}$C is predicted above the $\alpha+{}^{12}{\rm Be}$ and ${}^{6}{\rm He}+{}^{10}{\rm Be}$ thresholds, the decay pattern should be important information to identify the linear-chain.

In this work, based on AMD calculation, we study the positive- and negative-parity linear-chain
states of $^{16}{\rm C}$ and  discuss their decay patterns. 
The positive-parity linear-chain has the valence neutrons occupying molecular-orbits $(3/2^-_\pi)^2(1/2^-_\sigma)^2$.
We predict that the linear-chain states primary decay to $^{12}$Be($2^+_1$) as well as to the $^{12}$Be(g.s.).
They will also decay to $^{10}$Be(g.s.) and $^{10}$Be($2^+_1$) by the ${}^{6}{\rm He}$ emission, which is a signature of the covalency of valence neutrons.

In the negative parity, two rotational bands composed of the linear-chain configuration are found.
One has the valence neutrons occupying the molecular-orbits $(3/2^-_\pi)^2(1/2^-_\sigma)(3/2^+_\pi)$, and the other does not have the clear molecular-orbits configuration.
Their $\alpha$- and ${}^{6}{\rm He}$ reduced widths are smaller than those of positive-parity linear-chain band, but sufficiently large to be distinguished from other non-cluster states.

The paper is organized as follows.
The AMD framework is briefly explained in the next section.
In Sec. III, the density distribution on the energy surface, excitation energies and decay widths  are discussed for positive- and negative-parity.
In the last section, we summarize this work.

\section{theoretical framework}

\subsection{variational calculation and generator coordinate method}
The microscopic $A$-body Hamiltonian used in this work reads
\begin{align}
 \hat{H} = \sum_{i=1}^A \hat{t}_i - \hat{t}_{c.m.} + \sum_{i<j}^A \hat{v}^N_{ij} + \sum_{i<j}^Z \hat{v}^C_{ij},
\end{align}
where the Gogny D1S interaction \cite{gogn91} is used as an effective nucleon-nucleon interaction $\hat{v}^N$.
It reproduces the $\alpha$- and $^6$He- threshold energies in $^{16}$C.
The Coulomb interaction $\hat{v}^C$ is approximated by a sum of seven Gaussians. 
The kinetic energy of the center-of-mass $\hat{t}_{c.m.}$ is exactly removed. 

The AMD intrinsic wave function $\Phi_{int}$ is represented by a Slater determinant of single
particle wave packets, 
\begin{align}
 \Phi_{int} ={\mathcal A} \{\varphi_1,\varphi_2,...,\varphi_A \}
 =\frac{1}{\sqrt{A!}}\mathrm{det}[\varphi_{i}({\bm r}_j)],
  \label{EQ_INTRINSIC_WF}  
\end{align}
where $\varphi_i$ is the single particle wave packet which is a direct product of the deformed
Gaussian spatial part \cite{kimu04}, spin ($\chi_i$) and isospin ($\xi_i$) parts,  
\begin{align}
 \varphi_i({\bm r}) &= \phi_i({\bm r})\otimes \chi_i\otimes \xi_i, \label{eq:singlewf}\\
 \phi_i({\bm r}) &= \exp\biggl\{-\sum_{\sigma=x,y,z}\nu_\sigma\Bigl(r_\sigma -\frac{Z_{i\sigma}}{\sqrt{\nu_\sigma}}\Bigr)^2\biggr\}, \\
 \chi_i &= a_i\chi_\uparrow + b_i\chi_\downarrow,\quad
 \xi_i = {\rm proton} \quad {\rm or} \quad {\rm neutron}.\nonumber
\end{align}
The centroids of the Gaussian wave packets $\bm Z_i$, the direction of nucleon spin $a_i, b_i$,
and the width parameter of the deformed Gaussian $\nu_\sigma$ are the variational parameters.  
The intrinsic wave function is projected to the eigenstate of the parity to investigate both of
the positive- and negative-parity states,
\begin{align}
 \Phi^\Pi &= P^\Pi\Phi_{int}=\frac{1+\Pi P_x}{2}\Phi_{int}, \quad \Pi=\pm, 
\end{align}
where $P^\Pi$ and $P_x$ denote parity projector and operator. Using this wave function, the
variational energy is defined as, 
\begin{align}
 E^\Pi = \frac{\braket{\Phi^\Pi|H|\Phi^\Pi}}{\braket{\Phi^\Pi|\Phi^\Pi}} 
\end{align}
By the frictional cooling method \cite{enyo95}, the variational parameters are determined so that $E^\Pi$ is minimized.
In this study, we add the constraint potential to the variational energy,
\begin{align}
 {\tilde E}^\Pi = \frac{\braket{\Phi^\Pi|H|\Phi^\Pi}}{\braket{\Phi^\Pi|\Phi^\Pi}} 
 + v_\beta(\braket{\beta} - \beta)^2  + v_\gamma (\braket{\gamma} - \gamma)^2,
\end{align}
where $\braket{\beta}$ and $\braket{\gamma}$ are the quadrupole deformation parameters of the intrinsic wave function defined in Ref. \cite{suha10,kimu12}, and $v_\beta$ and $v_\gamma$ are chosen large enough that $\braket{\beta}$ and $\braket{\gamma}$ are equal to $\beta$ and $\gamma$ after the variation.
By minimizing ${\tilde E}^\Pi$, we obtain the optimized wave function $\Phi^\Pi(\beta,\gamma)=P^\Pi\Phi_{int}(\beta,\gamma)$ which has the minimum energy for each set of $\beta$ and $\gamma$.
It is noted that our previous work employed only $\beta$-constraint, therefore the degree-of-freedom of $\gamma$ deformation was not explicitly included. 

After the variational calculation, the eigenstate of the total angular momentum $J$ is projected
out from  $\Phi^\Pi(\beta,\gamma)$,
\begin{align}
 \Phi^{J^\Pi}_{MK}(\beta,\gamma) &=  P^{J}_{MK}\Phi^\Pi(\beta,\gamma)\nonumber\\
  &=\frac{2J+1}{8\pi^2}
  \int d\Omega D^{J*}_{MK}(\Omega)\hat{R}(\Omega)\Phi^{\Pi}(\beta,\gamma).
\end{align} 
Here, $P^{J}_{MK}$, $D^{J}_{MK}(\Omega)$  and $\hat{R}(\Omega)$ are the angular momentum
projector, the Wigner $D$ function and the rotation operator, respectively. The integrals over
Euler angles $\Omega$ are evaluated numerically. 

Then, we perform the generator coordinate method (GCM) calculation by employing the quadrupole deformation parameters $\beta$
and $\gamma$ as the generator coordinates.  The wave function of GCM reads,

\begin{align}
 \Psi^{J^\Pi}_{Mn} = \sum_i\sum_Kc^{J^\Pi}_{Kin}\Phi^{J^\Pi}_{MK}(\beta_i,\gamma_i),\label{eq:gcmwf}
\end{align}
where the coefficients $c^{J^\Pi}_{Kin}$ and eigenenergies $E^{J^\Pi}_n$ are obtained by solving the
Hill-Wheeler equation \cite{hill54}, 
\begin{align}
 \sum_{i'K'}{H^{J^\Pi}_{KiK'i'}c^{J}_{K'i'n}} &= 
 E^{J^\Pi}_n \sum_{i'K'}{N^{J^\Pi}_{KiK'i'}c^{J^\Pi}_{K'i'n}},\\
  H^{J^\Pi}_{KiK'i'} &= \braket{\Phi^{J^\Pi}_{MK}(\beta_i,\gamma_i)|\hat{H}|
 \Phi^{J^\Pi}_{MK'}(\beta_{i'},\gamma_{i'})}, \nonumber\\
  N^{J^\Pi}_{KiK'i'} &= \braket{\Phi^{J^\Pi}_{MK}(\beta_i,\gamma_i)|
 \Phi^{J^\Pi}_{MK'}(\beta_{i'},\gamma_{i'})}.\nonumber
\end{align}
We also calculate the overlap between $\Psi_{Mn}^{J^\Pi}$ and the basis wave function of the GCM
$\Phi^{J^\Pi}_{MK}(\beta_i,\gamma_i)$, 
\begin{align}
 |\braket{\Phi^{J^\Pi}_{MK}(\beta,\gamma)|\Psi^{J^\Pi}_{Mn}}|^2/
 \braket{\Phi^{J^\Pi}_{MK}(\beta,\gamma)|\Phi^{J^\Pi}_{MK}(\beta,\gamma)},
\end{align}
to discuss the dominant configuration in each state described by $\Psi_{Mn}^{J^\Pi}$.

\subsection{single particle orbits}
The neutron single-particle orbits of the intrinsic wave functions $\Phi_{int}(\beta,\gamma)$ provide us the motion of the valence neutrons around the core nucleus.
In order to construct a single-particle Hamiltonian, we first transform the single particle wave packet $\varphi_i$ to the orthonormalized basis,
\begin{align}
 \widetilde{\varphi}_\alpha = \frac{1}{\sqrt{\lambda_\alpha}}\sum_{i=1}^{A}g_{i\alpha}\varphi_i.  
\end{align}
Here, $\lambda_\alpha$ and $g_{i\alpha}$ are the eigenvalues and eigenvectors of the
overlap matrix $B_{ij}=\langle\varphi_i|\varphi_j\rangle$.
Using this basis, the Hartree-Fock single particle Hamiltonian is derived,
\begin{align}
 h_{\alpha\beta} &=
  \langle\widetilde{\varphi}_\alpha|\hat{t}|\widetilde{\varphi}_\beta\rangle + 
  \sum_{\gamma=1}^{A}\langle
  \widetilde{\varphi}_\alpha\widetilde{\varphi}_\gamma|
  {\hat{v}^N+\hat{v}^C}|
  \widetilde{\varphi}_\beta
\widetilde{\varphi}_\gamma -
\widetilde{\varphi}_\gamma\widetilde{\varphi}_\beta\rangle \nonumber\\ 
 &+\frac{1}{2}\sum_{\gamma,\delta=1}^{A}
 \langle\widetilde{\varphi}_\gamma\widetilde{\varphi}_\delta 
|\widetilde{\varphi}_\alpha^*\widetilde{\varphi}_\beta
\frac{\delta\hat{v}^N}{\delta \rho}|\widetilde{\varphi}_\gamma
\widetilde{\varphi}_\delta - \widetilde{\varphi}_\delta  \widetilde{\varphi}_\gamma
\rangle.
\end{align}
The eigenvalues $\epsilon_s$ and eigenvectors  $f_{\alpha s}$ of $h_{\alpha\beta}$ give the single particle energies and the single particle orbits,
$\widetilde{\phi}_s = \sum_{\alpha=1}^{A}f_{\alpha s}\widetilde{\varphi}_\alpha$.
We calculate the amount of the positive-parity component in the single-particle orbit,  
\begin{align}
 p^+ = |\langle \widetilde{\phi}_s|\frac{1+P_x}{2}| \widetilde{\phi}_s\rangle|^2, \label{eq:sp1}
\end{align}
and angular momenta in the intrinsic frame,
\begin{align}
 j(j+1)&= \langle \widetilde{\phi}_s|\hat{j}^2| \widetilde{\phi}_s\rangle, \quad
 |j_z| = \sqrt{\langle \widetilde{\phi}_s|\hat{j}_z^2| \widetilde{\phi}_s\rangle},\label{eq:sp2}\\
 l(l+1)&= \langle \widetilde{\phi}_s|\hat{l}^2| \widetilde{\phi}_s\rangle, \quad
 |l_z| = \sqrt{\langle \widetilde{\phi}_s|\hat{l}_z^2| \widetilde{\phi}_s\rangle},\label{eq:sp3}
\end{align}
which are used to discuss the properties of the single particle orbits.

\subsection{reduced width amplitude and decay width}
Using the GCM wave function, we calculate the reduced width amplitudes (RWA)
$y_{lj^{\pi}_n}(r)$ for the $\alpha+{}^{12}{\rm Be}$ and $^6{\rm He}+{}^{10}{\rm Be}$ decays which are defined as,
\begin{align}
 y_{lj^{\pi}_n}(r) = \sqrt{\frac{A!}{A_{\rm He}!A_{\rm Be}!}}
 \langle \phi_{\rm He}[\phi_{\rm Be}(j^{\pi}_n)Y_{l0}({\hat r})]_{J^\Pi M}
 |\Psi^{J\Pi}_{Mn}\rangle,\label{eq:rwa}
\end{align}
where $\phi_{\rm He}$ denotes the ground state wave function for $^4{\rm He}$ or $^6{\rm He}$, and $\phi_{\rm Be}(j^{\pi}_n)$ denotes the wave functions for daughter nucleus $^{12}{\rm Be}$ or $^{10}{\rm Be}$ with spin-parity $j^{\pi}_n$.
$Y_{l0}({\hat r})$ is the orbital angular momentum of the inter-cluster motion, and it is coupled with the angular momentum of Be$(j^{\pi}_n)$ to yield the total spin-parity $J^\Pi$.
$A_{\rm He}$ and $A_{\rm Be}$ are the mass numbers of He and Be, respectively.
The reduced width $\gamma_{lj^{\pi}_n}$ is given by the square of the RWA,
\begin{align}
 \gamma^2_{lj^{\pi}_n}(a) = \frac{\hbar^2}{2\mu a}|ay_{lj^{\pi}_n}(a)|^2,
\end{align}
and the spectroscopic factor $S$ is defined by the integral of the RWA,
\begin{align}
 S = \int^{\infty}_0 r^2|y_{lj^{\pi}_n}(r)|^2dr.
\end{align}
The partial decay width is a product of the reduced width and the penetration factor $P_l(a)$,
\begin{align}
 \Gamma_{lj^{\pi}_n} &= 2P_l(a)\gamma^2_{lj^{\pi}_n}(a), \quad
 P_l(a) = \frac{ka}{F^2_l(ka)+G^2_l(ka)}, 
\end{align}
where $a$ denote the channel radius, and $P_l$ is given by the Coulomb
regular and irregular wave functions $F_l$ and $G_l$. The wave number $k$ is determined by the decay
$Q$-value and the reduced mass $\mu$ as $k=\sqrt{2\mu E_Q}$.

In order to calculate the RWA, we employ the Laplace expansion method given in Ref. \cite{chib17}.
This method is applicable to unequal-sized and deformed clusters without any approximation.
%The RWA is obtained by using the GCM wave function,
%\begin{align}
% y_{lj^{\pi}_n}(a) &= \frac{1}{\sqrt{n^{J\Pi}_K}}
% \sum_{1\leq j_1<...<j_A\leq A}P(i_1,...,i_{C_1})\\
% &\times \bigl[ \chi_{l} (a;i_1,...,i_A)[N^{j_1\pi_1}(i_1,...,i_{C_1})\\
% &\quad \times N^{j_2\pi_2}(i_{C_1+1},...,i_A)]_{j_{12}}\bigl]_{JK}, \label{eq:laprwa}
%\end{align}
%with the definitions of the overlaps:
%\begin{align}
% \chi_{l} (a;i_1,...,i_A) &= \biggl\langle \frac{\delta(r-a)}{r^2}Y_{l0}(\hat{r})\biggl| \chi({\bm r};i_1,...,i_A)\biggl\rangle,\\
% N^{j_1\pi_1}_{m_1}(i_1,...,i_{C_1}) &= \langle\Phi^{j_1\pi_1}_{m_1He}|\Psi_{He}(i_1,...,i_{C_1})\rangle,\\
% N^{j_2\pi_2}_{m_2}(i_{C_1+1},...,i_A) &= \langle\Phi^{j_2\pi_2}_{m_2Be}|\Psi_{Be}(i_1,...,i_{C_1})\rangle,
%\end{align}
%where $P(i_1,...,i_{C_1})$ and $n^{J\Pi}_K$ denotes the phase factor and normalization factor, respectively.
The intrinsic wave functions for $^{10,12}$Be and $^{4,6}$He are generated by the AMD energy variation.
For $^{10}$Be, we obtained two different intrinsic wave functions in which two valence neutrons occupy so-called $\pi$- and $\sigma$-orbits, respectively.
We regard that the former correspond to the ground band (the $0^+_1$ and $2^+_1$ states), while the latter is the excited band (the $0^+_2$ and $2^+_3$ states).
For $^{12}{\rm Be}$, we obtain an intrinsic wave function in which two of four valence neutrons occupy $\pi$-orbit and the others occupy $\sigma$-orbit which is regarded as the $^{12}$Be$(0^+_1)$ and $^{12}$Be$(2^+_1)$.
We also obtained another configuration having four valence neutrons in $\pi$-orbit, which we regard as $^{12}$Be$(0^+_2)$ and $^{12}$Be$(2^+_2)$.
However, we found that the decay width to the these states are negligibly small, and hence, they are not discussed here. 
In following calculation, we assume that the $^{4}$He and $^{6}$He clusters are always $j^\pi=0^+$. 

We also calculate the neutron spectroscopic factors in order to compare with the $\alpha$-cluster spectroscopic factors.
The neutron spectroscopic factor $S_n$ reads,
\begin{align}
 S_n = \int^{\infty}_0 r^2|\varphi(r)|^2dr,
\end{align}
where $\varphi(r)$ is the overlap amplitude which is the overlap between the wave functions of nuclei with mass $A$ and $A+1$,
\begin{align}
 \varphi(r) = \sqrt{A+1}\braket{\Psi^{J'^{\Pi\prime}}_{M'n'}(^{15}{\rm C})|\Psi^{J^\Pi}_{Mn}(^{16}{\rm C})}.
\end{align}
The intrinsic wave function for $^{15}$C is generated by the AMD energy variation.
The spin-parity of $^{15}$C are chosen as $J'^{\Pi\prime} = 1/2^+$ (the ground state of $^{15}$C) for positive-parity states of $^{16}$C and $J'^{\Pi\prime} = 1/2^-, 3/2^-, 5/2^-$ for negative-parity states.

\section{Results and Discussion}
%In Sec. III A, we summarize the properties of the $\pi$-bond and $\sigma$-bond linear chains studied in our previous work \cite{baba16}.
%In Sec. III B and C, by referring the latest experimental data and the theoretical analysis of the decay modes, we discuss the assignment of the linear-chain bands.

\subsection{Energy surface and intrinsic structures}
\begin{figure}[h]
 \centering
 \includegraphics[width=1.0\hsize]{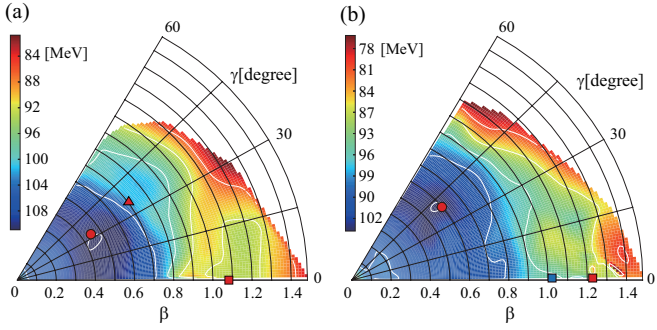}
 \caption{(color online) The angular momentum projected energy surfaces for (a) the $J^\pi=0^+$ and (b) $J^\pi=1^-$ states as functions of quadrupole deformation parameters $\beta$ and $\gamma$. 
 The filled circles, triangles and boxes in the panel (a) show the ground, triangular and linear-chain structures, while in the panel (b), the circle shows the position of the energy minimum and filled boxes show the linear-chain configurations.} 
 \label{fig:surface}
\end{figure}

\begin{figure}[h]
 \centering
 \includegraphics[width=1.0\hsize]{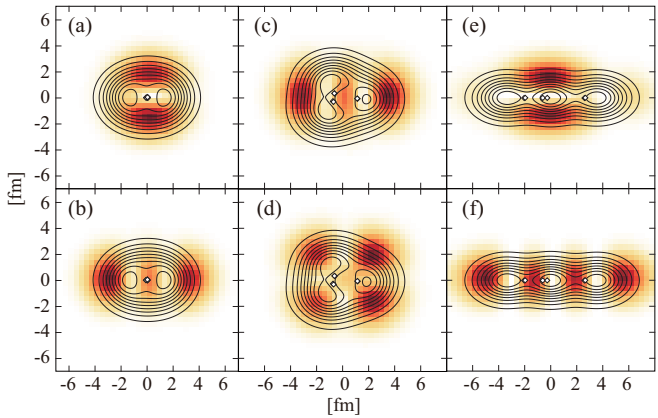}
 \caption{(color online) The density distributions of positive parity states of the ground (a)(b), triangular (c)(d), and linear-chain (e)(f) configurations.
 The contour lines show the proton density distributions.
 The color plots show the single particle orbits occupied by four valence neutrons.
 The lower panels show the most weakly bound two neutrons, while the upper panels show the other two valence neutrons.
 Open boxes show the centroids of the Gaussian wave packets describing protons.} 
 \label{fig:density+}
\end{figure}

\begin{table}[h]
\caption{The properties of the valence neutron orbit shown in Fig. \ref{fig:density+}. 
 Each column shows the single particle energy $\varepsilon$ in MeV, the amount of the
 positive-parity component $p^+$ and the angular momenta defined by
 Eqs. (\ref{eq:sp1})-(\ref{eq:sp3}).} 
\label{tab:spo+}
\begin{center}
 \begin{ruledtabular}
  \begin{tabular}{ccccccc} 
    & $\varepsilon $ &$p^+$ & $j$ & $|j_{z}|$ & $l$ & $|l_{z}|$ \\ \hline
	(a) & -8.69 & 0.01 & 0.7 & 0.5 & 1.1 & 1.0 \\
	(b) & -3.95 & 0.99 & 2.2 & 0.5 & 1.8 & 0.4 \\ 
	(c) & -5.84 & 0.98 & 2.2 & 1.9 & 1.9 & 1.6 \\
	(d) & -2.97 & 0.98 & 2.4 & 1.9 & 2.1 & 1.8 \\
	(e) & -6.31 & 0.05 & 1.8 & 1.4 & 1.4 & 1.0 \\
	(f) & -3.16 & 0.07 & 2.8 & 0.6 & 2.6 & 0.3 \\
  \end{tabular}
  \end{ruledtabular}
\end{center}
\end{table}

\begin{figure}[h]
 \centering
 \includegraphics[width=1.0\hsize]{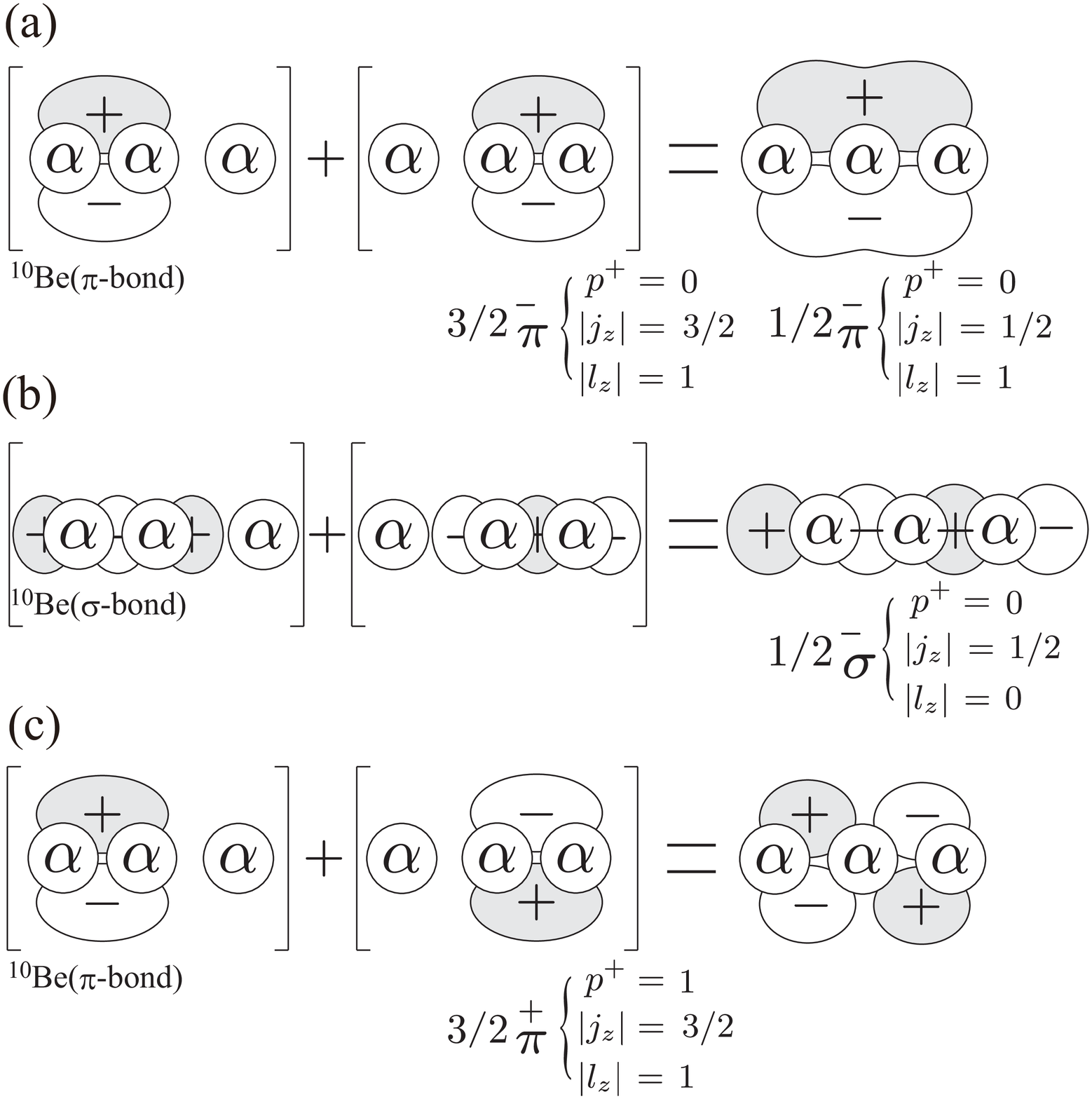}
 \caption{The schematic figure showing the $\pi$ and $\sigma$-orbits around the linear chain. 
 The combination of the $\pi$-orbits around $^{10}$Be perpendicular to the symmetry
 axis generates $\pi$-orbits, while the combination of parallel orbits
 around $^{10}$Be generates $\sigma$-orbit.} 
 \label{fig:mol}
\end{figure}

In Fig. \ref{fig:surface}, the energy surfaces for $J^\pi=0^+$ and $J^\pi=1^-$ states are shown as the function of quadrupole deformation parameters $\beta$ and $\gamma$.
The circles on the energy surfaces show the position of the energy minima.
First, we discuss three different structures on the energy surface of positive parity based on their intrinsic density distributions shown in Fig. \ref{fig:density+}.
Although the $\beta\gamma$-constrained AMD method was newly applied in this study, these three structures are almost identical to those discussed in our previous work \cite{baba14}.

The energy minimum of the $0^+$ state is located at $(\beta,\gamma)=(0.45,31^\circ)$ with the binding energy of 110.5 MeV.
The intrinsic density distribution at the minimum is shown in Figs. \ref{fig:density+} (a)(b).
As clearly seen, this structure has no outstanding clustering.
Figs. \ref{fig:density+} (c)(d) show a different structure which we call triangular configuration located around $(\beta,\gamma)=(0.70,37^\circ)$.
The 3$\alpha$ cluster core forms triangle configuration as seen in the density distribution.
Table. \ref{tab:spo+} (c)(d) show that four valence neutrons occupy the $(sd)^4$ shell, indicating $2\hbar\omega$ excitation.
However, due to its asymmetric shape, the valence proton orbits are an admixture of the positive- and negative-parity components.
We note that a similar configuration appears in $^{14}$C, which also have valence neutrons shown in the panel (c) but without those shown in the panel (d).

Further increase of the deformation realizes the linear-chain configuration in the strongly prolate deformed region.
In this region, there is an energy plateau around the local energy minimum located at $(\beta,\gamma)=(1.08,0^\circ)$.
As seen in Fig. \ref{fig:density+} (e)(f), its proton density distribution shows striking 3$\alpha$ cluster configuration with linear alignment.
In addition, the properties of valence neutron orbits listed in Table. \ref{tab:spo+} (e)(f) show that two valence neutrons occupy the so-called $\pi$-orbit and the other two neutrons occupy the $\sigma$-orbit.
Here the $\pi$-orbit in $^{16}$C is formed by the in-phase linear-combination of the $\pi$-orbit of $^{10}$Be and denoted as $3/2^{-}_\pi$ and $1/2^{-}_\pi$ depending on the value of $|j_z|$ as illustrated in Fig. \ref{fig:mol} (a).
The $\sigma$-orbit is a linear-combination of $\sigma$-orbit of $^{10}$Be as illustrated in Fig. \ref{fig:mol} (b).
Therefore, with these definitions, this state is regarded to have the $(3/2^{-}_\pi)^2$$(1/2^{-}_\sigma)^2$ configuration.

The energy minimum of the energy surface for the $1^-$ states (Fig. \ref{fig:surface} (b)) is located at $(\beta,\gamma)=(0.59,40^\circ)$ with the binding energy of $103.6$ MeV.
At the minimum, the single-particle properties show the $1p1h$-configuration $\nu (p_{1/2})^{-1}(d_{5/2})^1$.

\begin{figure}[h]
 \centering
 \includegraphics[width=1.0\hsize]{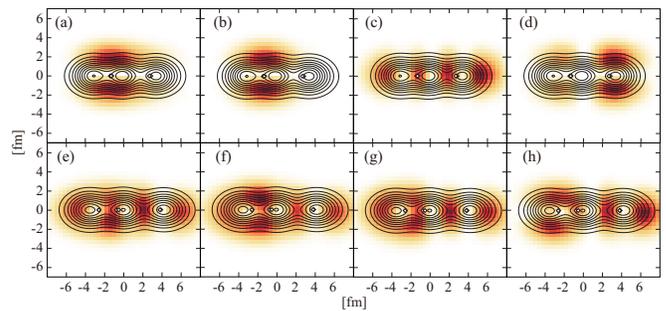}
 \caption{(color online) The density distributions of negative parity linear-chain states.
 The panels (a)-(d) correspond the state at $(\beta,\gamma)=(1.02,1^\circ)$,
 while the panels (e)-(h) correspond the state at $(\beta,\gamma)=(1.23,1^\circ)$.
 The contour lines show the proton density distributions.
 The color plots show the single particle orbits occupied by four valence neutrons.
 The panels (a) and (e) show the most deeply bound valence neutrons,
 while the panels (d) and (h) show the most weakly bound neutrons.
 Open boxes show the centroids of the Gaussian wave packets describing protons.} 
 \label{fig:density-}
\end{figure}

Figs. \ref{fig:density-} (a)-(d) show a basis wave function located at $(\beta,\gamma)=(1.02,1^\circ)$ in Fig. \ref{fig:surface}(b).
This linear-chain configuration appears in the prolate deformed region,  although there is no plateau in the energy surface of negative parity.
The density distribution and properties of valence neutron orbits show that the 3$\alpha$ core are linearly aligned and three valence neutrons (Figs. \ref{fig:density-} (a)-(c)) occupy $(3/2^{-}_\pi)^2$$(1/2^{-}_\sigma)^1$ orbits similar to the linear-chain configuration of positive parity.
However, the most weakly bound valence neutron (Fig. \ref{fig:density-} (d)) occupies a different orbit.
The properties of single particle orbit (Table. \ref{tab:spo-} (d)) show that the most weakly bound valence neutron occupies the {\it ungerade} $\pi$-orbit which is a linear-combination of $^{10}{\rm Be}$ $\pi$-orbit with anti-phase as illustrated in Fig. \ref{fig:mol} (c).
In addition, it can be seen that this $(3/2^{+}_\pi)$ orbit locates around $\alpha$ particle of right side preferably.
It is noted that this type of the linear-chain was not found in $^{14}{\rm C}$.
We consider that this orbit is unbound in $^{14}{\rm C}$, and makes the negative-parity linear-chain unstable in $^{14}{\rm C}$.
%Therefore, the linear-chain of negative parity may show the $^{5}{\rm He}+{}^{11}{\rm Be}$ configuration, which is consistent with their small reduced widths discussed later.

Figs. \ref{fig:density-} (e)-(h) show another intrinsic wave function belonging the linear-chain configuration appeared around $(\beta,\gamma)=(1.23,1^\circ)$.
Although the density distribution and properties of valence neutron orbit do not show the clear molecular orbit nature because of the strong parity mixing, this intrinsic wave function has the largest overlap with member states of a linear-chain band as mentioned in next section.

\begin{table}[h]
\caption{The properties of the valence neutron orbits shown in Fig. \ref{fig:density-}. 
 Each column shows the single particle energy $\varepsilon$ in MeV, the amount of the
 positive-parity component $p^+$ and the angular momenta defined by
 Eqs. (\ref{eq:sp1})-(\ref{eq:sp3}).} 
\label{tab:spo-}
\begin{center}
 \begin{ruledtabular}
  \begin{tabular}{ccccccc} 
   & $\varepsilon $ &$p^+$ & $j$ & $|j_{z}|$ & $l$ & $|l_{z}|$ \\ \hline
	(a) & -6.76 & 0.05 & 2.1 & 1.5 & 1.7 & 1.0 \\
	(b) & -6.63 & 0.16 & 2.0 & 1.5 & 1.7 & 1.0 \\ 
	(c) & -2.10 & 0.03 & 2.7 & 0.5 & 2.5 & 0.3 \\
	(d) & -0.78 & 0.92 & 3.0 & 1.5 & 2.7 & 1.0 \\ \hline
	(e) & -4.80 & 0.08 & 2.6 & 1.1 & 2.3 & 0.7 \\
	(f) & -4.48 & 0.16 & 2.7 & 1.2 & 2.4 & 0.8 \\
	(g) & -4.07 & 0.23 & 2.8 & 1.0 & 2.6 & 0.6 \\
	(h) & -2.23 & 0.57 & 3.0 & 1.2 & 2.8 & 0.8 \\
  \end{tabular}
  \end{ruledtabular}
\end{center}
\end{table}

\subsection{Excitation spectrum}
\begin{figure*}[t]
 \includegraphics[width=0.9\hsize]{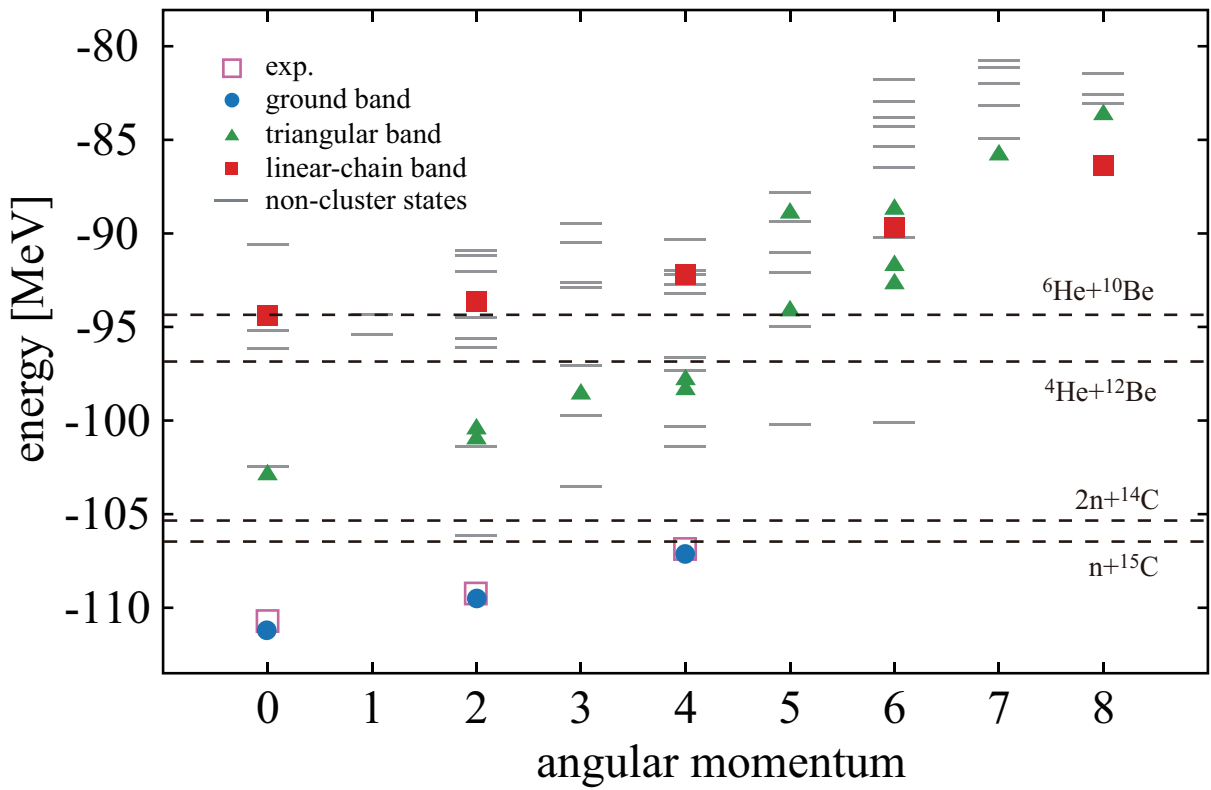}
 \caption{(color online) The positive-parity energy levels up  to $J^\pi=8^+$.
 Open boxes show the observed states with the definite spin-parity assignments, and other symbols show the calculated result.
 The filled circles, triangles and boxes show the ground, triangular and linear-chain bands, while lines show the non-cluster states which have the reduced widths smaller than 0.10 MeV$^{1/2}$ except for the triangular band.}
 \label{fig:spectrum+} 
\end{figure*}

\begin{table}[h]
 \caption{Excitation energies (MeV), $\alpha$ reduced widths (MeV$^{1/2}$), $\alpha$-cluster and neutron spectroscopic factors of several selected positive-parity states. The reduced widths, $\alpha$ and neutron S-factors are calculated for the decays to the ground states of daughter nuclei.}
\label{tab:values+}
\begin{center}
 \begin{ruledtabular}
  \begin{tabular}{lccccc} 
   band & $J^\pi$ & $E_x$ & $\gamma_\alpha(6.0 {\rm fm})$ & $S_{\alpha}$ & $S_{n}$\\
   \hline
   ground & $0^+_1$ & 0.00 & 0.00 & 0.03 & 0.22 \\
               & $2^+_1$ & 1.69  & 0.00 & 0.00 & 0.35 \\
               & $4^+_1$ & 4.04 & 0.00 & 0.00 & 0.01 \\
   \hline
   triangular & $0^+_2$ & 8.35  & 0.01 & 0.05 & 0.12 \\
                 & $2^+_4$ & 10.22 & 0.00 & 0.00 & 0.01 \\
                 & $2^+_5$ & 10.79 & 0.00 & 0.01 & 0.02 \\
   \hline
   linear-chain & $0^+_6$ & 16.81 & 0.28 & 0.11 & 0.00 \\
                    & $2^+_{9}$ & 17.51 & 0.23 & 0.07 & 0.00 \\
                     & $4^+_{10}$ & 18.99 & 0.26 & 0.09 & 0.00 \\
                     & $6^+_5$ & 21.49 & 0.23 & 0.07 & 0.00 \\
  \end{tabular}
 \end{ruledtabular}
 \end{center}
\end{table}

\begin{table}[h!]
 \caption{The calculated in-band $B(E2)$ strengths for the low-spin positive-parity states in unit of  $e^2\rm fm^4$. 
 The number in parenthesis is the observed data \cite{imai04,ong06,ong08,wied08,petr12}. } 
\label{tab:be2+}
\begin{center}
 \begin{ruledtabular}
 \begin{tabular}{lcc}
   & $J_i\rightarrow J_f$ & $B(E2;J_i\rightarrow J_f)$ \\ \hline
  ground $\rightarrow$ ground& $2^{+}_{1}\rightarrow 0^{+}_{1}$ & 6.7 (0.92$\sim$4.2)\\ 
                                        & $4^{+}_{1}\rightarrow 2^{+}_{1}$ & 4.1 \\\hline
  triangular & $2^{+}_{4} \rightarrow 0^{+}_{2}$ & 2.5 \\
  $\rightarrow$ triangular & $2^{+}_{5}\rightarrow 0^{+}_{2}$ & 0.9 \\
                                        & $3^{+}_{3}\rightarrow 2^{+}_{4}$ & 9.5 \\
                                        & $3^{+}_{3}\rightarrow 2^{+}_{5}$ & 8.5 \\\hline
  linear-chain & $2^{+}_{9}\rightarrow 0^{+}_{6}$ & 380.3\\
  $\rightarrow$ linear-chain & $4^{+}_{10}\rightarrow 2^{+}_{9}$ & 544.3\\
                                          & $6^{+}_{5}\rightarrow 4^{+}_{10}$ & 891.4\\
 \end{tabular}
 \end{ruledtabular}
\end{center}
\end{table}

Figure \ref{fig:spectrum+} shows the spectrum of the positive-parity states obtained by the GCM calculation.
The properties of the several selected states are listed in Tab. \ref{tab:values+}.
For the positive-parity, it is found that three different bands exist; ground, triangular and linear-chain bands.
We classified the excited states which have $\alpha$ reduced widths larger than 0.10 MeV$^{1/2}$ at the channel radius $a=6.0$ fm as cluster states.
In the present result, only the linear-chain band satisfies this condition.
For the triangular configuration, the member states have overlap larger than 0.50 with the configuration shown in Fig. \ref{fig:density+} (c)(d) are classified as the triangular band.
The intra-band $B(E2)$ strengths are listed in Tab. \ref{tab:be2+}.

The member states of the ground band are dominantly composed of the configurations around the energy minimum of the energy surface.
The ground state has the largest overlap with the basis wave function shown in Fig. \ref{fig:density+} (a)(b) that amounts to 0.98, and the calculated binding energy is $111.2$ MeV which is close to the observed binding energy of $110.8$ MeV.
The excitation energies of other member states $2^+_1$ and $4^+_1$ are also described well.
This band has no outstanding clustering but has a shell model like structure with a $\nu(sd)^2$ configuration which can be confirmed from the small $\alpha$ cluster spectroscopic factors and large neutron spectroscopic factors given in Tab. \ref{tab:values+}.

Because of its triaxial deformed shape, the triangular configuration generates two rotational bands built on the $0^+_2$ and $2^+_5$ states.
The member states have overlap larger than 0.50 with the configuration shown in Fig. \ref{fig:density+} (c)(d) which amount to, for example, 0.78 in the case of the $0^+_2$ state.
The member states with $J^\pi\geq 5^+$ are fragmented into several states because of  the coupling with the non-cluster configurations. 
Compared to the linear-chain states, these bands have less pronounced clustering and $\alpha$ clusters are considerably distorted.
As a result, the member states gain binding energy and the low-spin states locate well below the cluster thresholds.
Because of the $\alpha$-cluster distortion and deeper binding, the triangular configuration has small $\alpha$ spectroscopic factors and reduced widths as listed in Tab. \ref{tab:values+}.

The linear-chain configuration generates a rotational band which built on the $0^+_6$ state located at $16.7$ MeV.
The band-head $0^+_6$ state has the largest overlap with the configuration shown in Fig. \ref{fig:density+} (e)(f) that  amounts to 0.94.
The moment of inertia is estimated as $\hbar/2\Im=$112 keV which is considerably larger than those of ground band ($\hbar/2\Im=$196 keV) and triangular band ($\hbar/2\Im=$238 keV).
Owing to its large moment of inertia, the member state $J^\pi=8^+$ located at $E_x = 24.8$ MeV becomes the yrast state.
In addition, the large moment of inertia brings about the huge intra-band B(E2) compared with those of ground and triangular bands as listed in Tab. \ref{tab:be2+}.
In contrast to the ground and triangular bands, the linear-chain band has the large $\alpha$ cluster spectroscopic factors and very small neutron spectroscopic factors.
As all member states locate above the $^{4}{\rm He}+{}^{12}{\rm Be}$ and $^{6}{\rm He}+{}^{10}{\rm Be}$ thresholds, the linear-chain states should decay into these two channels, which can be an important observable to identify the linear-chain state as discussed in the next section.

\begin{figure*}[t]
 \includegraphics[width=0.9\hsize]{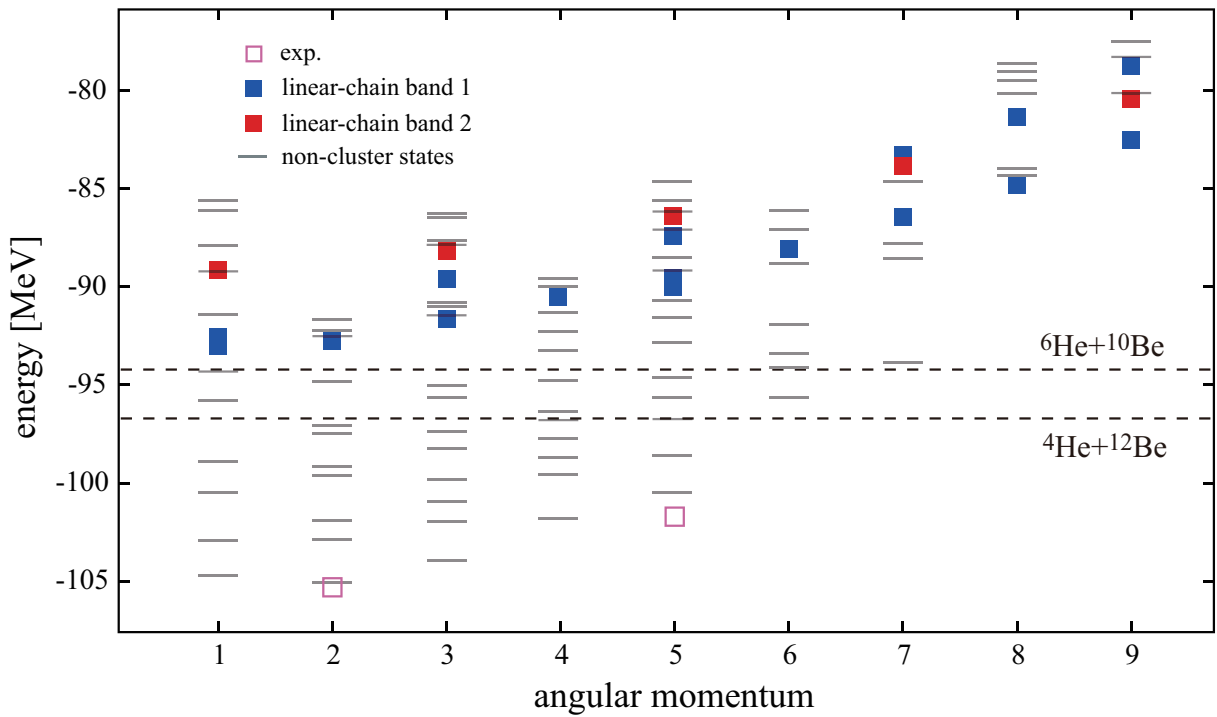}
 \caption{(color online) The negative-parity energy levels up to $J^\pi=9^-$. 
 Open boxes show the observed states with the definite spin-parity assignments \cite{bohl03,sato14}, and other symbols show the calculated result.
 The filled boxes show the linear-chain bands, while lines show the non-cluster states which have the reduced widths lower than 0.10 MeV$^{1/2}$.}
 \label{fig:spectrum-} 
\end{figure*}

Figure \ref{fig:spectrum-} shows the spectrum of the negative-parity states.
Only two states ($2^-$ and $5^-$), which are described by open boxes in the figure, were observed with the definite spin-parity assignments \cite{bohl03,sato14}.
Our calculation shows the yrast band which is built on the $2^-_1$ state located at $6.0$ MeV, and the $2^-_1$ and $5^-_1$ member states of this rotational band are close to the observed two states.
Since this band has the $1p1h$-configuration $\nu (p_{1/2})^{-1}(d_{5/2})^1$, the spectroscopic factors in the [$^{15}{\rm C}(g.s.)\otimes j$] channels are negligibly small but those in the [$^{15}{\rm C}(1/2^-)\otimes d_{5/2}$], [$^{15}{\rm C}(3/2^-)\otimes d_{5/2}$], and [$^{15}{\rm C}(5/2^-)\otimes d_{5/2}$] channels are large as listed in Tab. \ref{tab:sfactor}.

In the case of negative-parity, the linear-chain configuration generates two different types of rotational bands.
In the same manner to the positive-parity states, the excited states which have $\alpha$ reduced widths larger than 0.10 MeV$^{1/2}$ are classified as cluster states, and only linear-chain bands satisfy this condition.
These bands are also located above the ${}^{4}{\rm He}+{}^{12}{\rm Be}$ and $^{6}{\rm He}+{}^{10}{\rm Be}$ thresholds.
This is contrasting to $^{14}{\rm C}$ in which the linear-chain band was not obtained by the AMD calculations \cite{suha10,baba16,baba17}.
%In the case of the negative-parity linear-chain like configuration in $^{14}{\rm C}$, one proton in $\alpha$ core is excited, and hence, it cannot have $K=0$ quantum number.
The properties of the several selected linear-chain states are listed in Tab. \ref{tab:values-} and $B(E2)$ strengths are listed in Tab. \ref{tab:be2-}.

The first band, which we call the linear-chain band 1 (blue square), is dominantly composed of the wave function shown in Fig. \ref{fig:density-} (a)-(d) although it is mixed with non-cluster configurations.
Furthermore there is a mixing of $K=0^-$ and $1^-$ components.
As a result, the member states of this band is fragmented into several states.
For example, we classified both of the $1^-_7$ and $1^-_8$ states as the member states of the band, whose overlaps with the basis wave functions shown in Fig. \ref{fig:density-} (a)-(d) are 0.24 and 0.64, respectively.
%is built on the $1^-_7$ and $1^-_8$ states located around $E_x = 18.5$ MeV which is 1.7 MeV higher than the bandhead of the positive-parity linear-chain band.
%The bandheads are fragmented into two states due to the mixture with non-cluster states.
%The $1^-_7$ and $1^-_8$ state have the large overlap with the basis wave function shown in Fig. \ref{fig:density-} (a)-(d) which amounts to 0.24 and 0.64, respectively.
%The member states with higher angular momentum are fragmented into several states because they have $K = 1 or 2$ quantum numbers which are mixed.
Because of the fragmentation, the intra-band B(E2) values are smaller than those of positive-parity linear-chain band.
However, the moment of inertia, $\hbar/2\Im=$118 keV, is comparable with that of positive-parity linear-chain.
In this band, the neutron spectroscopic factors are negligible.
Compared with the positive-parity linear-chain, $\alpha$ spectroscopic factors are small, but sufficiently larger than other negative-parity states.

The other band, the linear-chain band 2 (red square), is built on the $1^-_{11}$ state at $E_x = 22.1$ MeV which is about 3.6 MeV higher than the linear-chain band 1.
The $1^-_{11}$ state has the largest overlap with the wave function shown in Fig. \ref{fig:density-} (e)-(h) which amounts to 0.92.
In contrast to the linear-chain band 1, the member states of this band have the $K=0$ quantum number and clearly form a single rotational band.
In addition, the moment of inertia, $\hbar/2\Im=$98 keV, and the intra-band B(E2) values are as large as those of positive-parity linear-chain band.
As well as the linear-chain band 1, the neutron spectroscopic factors are negligible while the $\alpha$ spectroscopic factors are a bit larger but smaller than those of positive-parity linear-chain band.

\begin{table}[h]
 \caption{Excitation energies (MeV), $\alpha$ reduced widths (MeV$^{1/2}$), $\alpha$-cluster  spectroscopic factors of several selected states for negative-parity. $\gamma_\alpha$ and $S_{\alpha}$ show the decay to the ground state ($0^+_1$) of $^{12}{\rm Be}$.}
\label{tab:values-}   
\begin{center}
 \begin{ruledtabular}
  \begin{tabular}{lccccc} 
   band & $J^\pi$ & $E_x$ & $\gamma_\alpha(5.5 {\rm fm})$ & $\gamma_\alpha(7.0 {\rm fm})$ & $S_{\alpha}$ \\
   \hline
   yrast band & $2^-_1$ & 6.11 & - & - & - \\
                    & $3^-_1$ & 7.25 & 0.00 & 0.00 & 0.02 \\
                    & $4^-_1$ & 9.34 & - & - & - \\
                    & $5^-_1$ & 10.71 & 0.00 & 0.00 & 0.00 \\
   \hline
   linear-chain & $1^-_7$ & 18.28 & 0.04 & 0.00 & 0.01 \\
   band 1        & $1^-_8$ & 18.64 & 0.02 & 0.01 & 0.00 \\
                    & $3^-_9$ & 19.45 & 0.10 & 0.01 & 0.03 \\
                    & $3^-_{13}$ & 21.57 & 0.05 & 0.02 & 0.01 \\
   \hline
   linear-chain & $1^-_{11}$ & 22.05 & 0.04 & 0.12 & 0.03 \\
   band 2         & $3^-_{14}$ & 23.00 & 0.04 & 0.12 & 0.03 \\
                     & $5^-_{15}$ & 24.76  & 0.03 & 0.11 & 0.02 \\
                     & $7^+_6$ & 27.35 & 0.06 & 0.11 & 0.01 \\
  \end{tabular}
 \end{ruledtabular}
 \end{center}
\end{table}

\begin{table}[h]
 \caption{Neutron spectroscopic factors of yrast band for negative-parity. The components of $^{15}{\rm C}\otimes s_{1/2}$ and $^{15}{\rm C}\otimes d_{3/2}$ are negligibly small.}
\label{tab:sfactor}
\begin{center}
 \begin{ruledtabular}
  \begin{tabular}{lcccc} 
    & $2^-_1$ & $3^-_1$ & $4^-_1$ & $5^-_1$ \\
   \hline
   $^{15}{\rm C}(1/2^-)\otimes d_{5/2}$ & 0.03 & 0.42 & - & - \\
   $^{15}{\rm C}(3/2^-)\otimes d_{5/2}$ & 0.34 & 0.04 & 0.19 & - \\
   $^{15}{\rm C}(5/2^-)\otimes d_{5/2}$ & 0.67 & 0.07 & 0.48 & 0.27 \\
  \end{tabular}
 \end{ruledtabular}
 \end{center}
\end{table}
 
\begin{table}[h!]
 \caption{The calculated in-band $B(E2)$ strengths for the low-spin negative-parity states in unit of  $e^2\rm fm^4$. } 
\label{tab:be2-}
\begin{center}
 \begin{ruledtabular}
 \begin{tabular}{lcc}
   & $J_i\rightarrow J_f$ & $B(E2;J_i\rightarrow J_f)$ \\ \hline
  linear-chain band 1 & $2^{-}_{9}\rightarrow 1^{-}_{7}$ & 53.2\\
                             & $2^{-}_{9}\rightarrow 1^{-}_{8}$ & 25.0\\
                             & $3^{-}_{9}\rightarrow 2^{-}_{9}$ & 37.7\\
                             & $3^{-}_{13}\rightarrow 2^{-}_{9}$ & 0.1\\
                             & $4^{-}_{11}\rightarrow 3^{-}_{9}$ & 72.0\\
                             & $4^{-}_{11}\rightarrow 3^{-}_{13}$ & 4.0\\ \hline
  linear-chain band 2 & $3^{-}_{14}\rightarrow 1^{-}_{11}$ & 492.2\\
                             & $5^{-}_{15}\rightarrow 3^{-}_{14}$ & 561.6\\
                             & $7^{-}_{6}\rightarrow 5^{-}_{15}$ & 556.1\\
 \end{tabular}
 \end{ruledtabular}
\end{center}
\end{table}

\subsection{Decay mode}
\begin{figure*}[h]
 \centering
 \includegraphics[width=0.9\hsize]{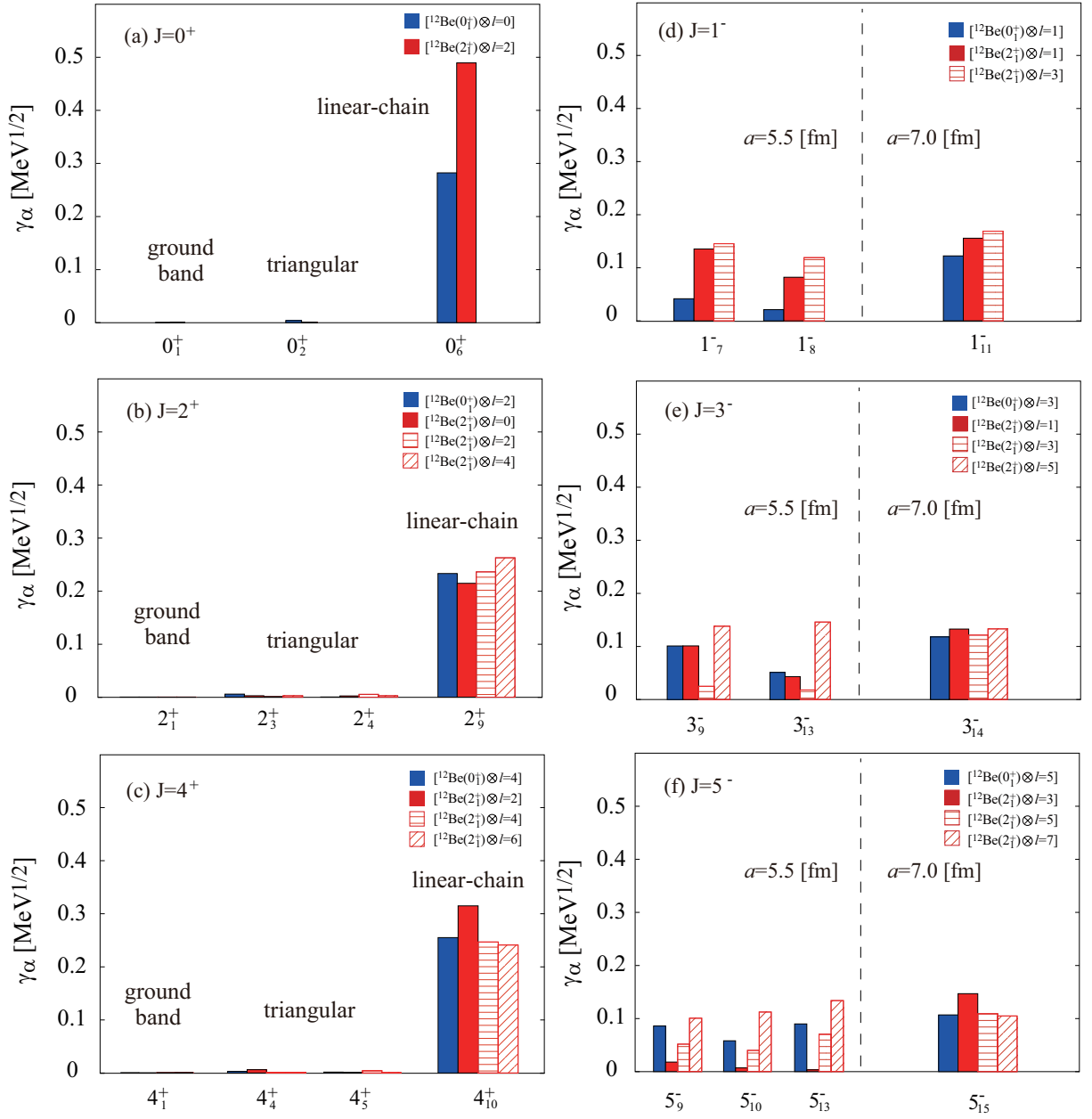}
 \caption{(color online) The calculated $\alpha$-decay reduced widths.
 Panels (a)-(c) show the decay of the  positive-parity states
 to the ground band of $^{12}{\rm Be}$.
 Panels (d)-(f) show the decay of the negative-parity states to the ground band of 
 $^{12}{\rm Be}$.
 The channel radii $a$ are 6.0 fm for (a)-(c) and 5.5(left side), 7.0(right side) fm for (d)-(f), respectively.}  
 \label{fig:width1} 
\end{figure*}

\begin{figure*}[h]
 \centering
 \includegraphics[width=0.9\hsize]{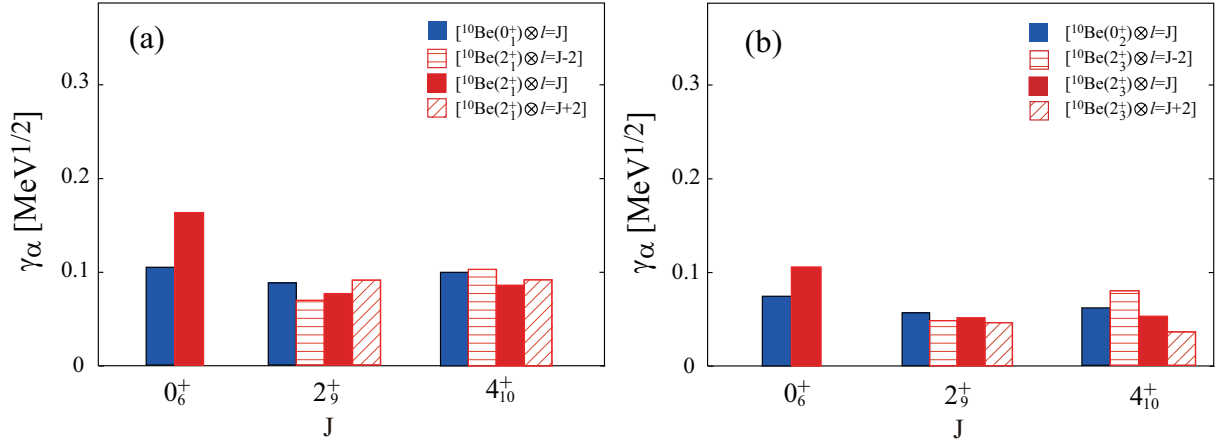}
 \caption{(color online) The calculated ${}^6{\rm He}$-decay reduced widths of linear-chain states in positive parity.
 In panel (a), the decay to the ground band of $^{10}{\rm Be}$ is shown.
 In panel (b), the decay to the excited band of $^{10}{\rm Be}$ is shown.
 The channel radius $a$ is 6.0 fm.}  
 \label{fig:width2+} 
\end{figure*}

\begin{figure*}[h]
 \centering
 \includegraphics[width=0.9\hsize]{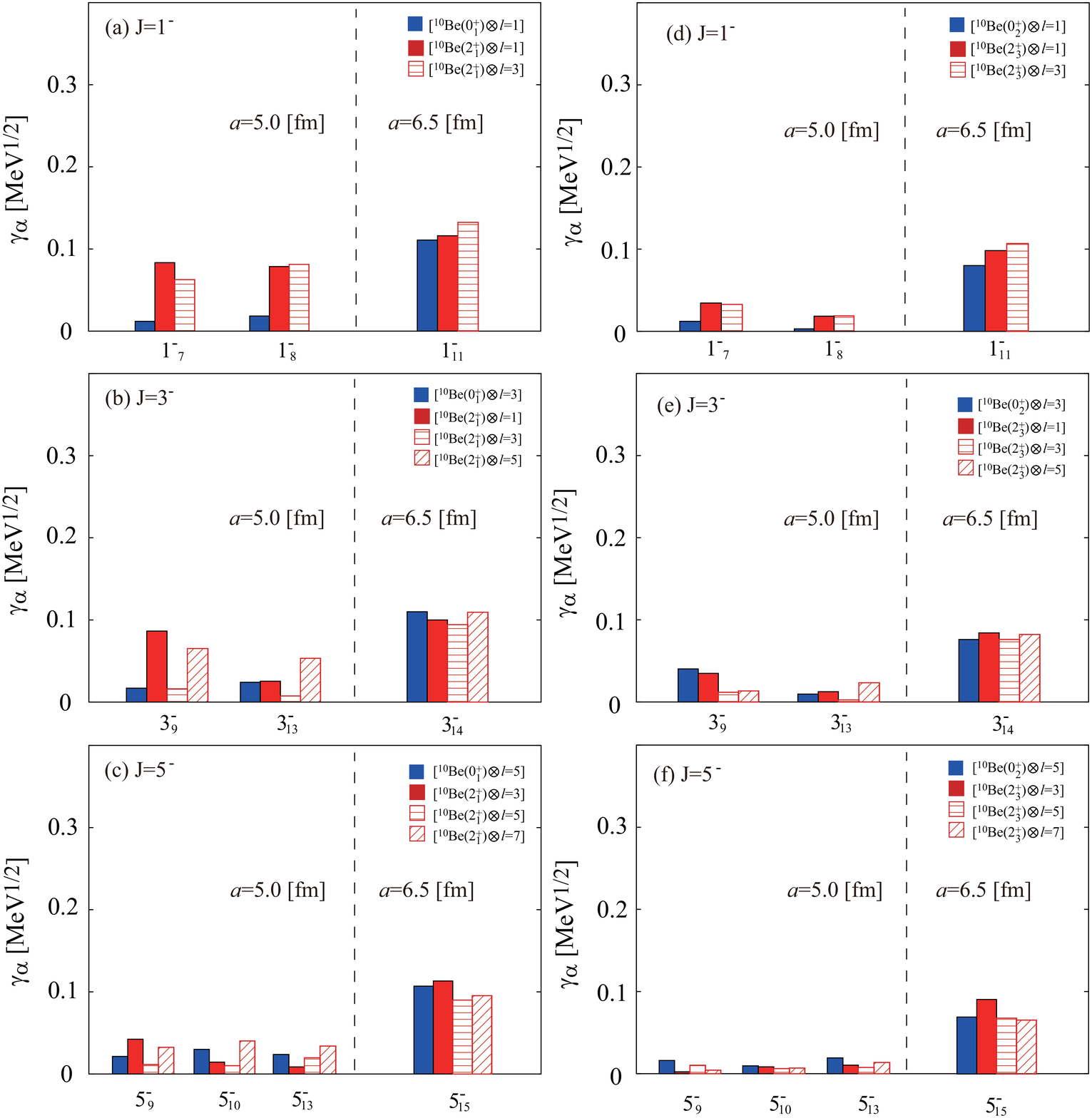}
 \caption{(color online) The calculated ${}^6{\rm He}$-decay reduced widths of linear-chain states in negative parity.
 Panels (a)-(c) show the decay to the ground band of $^{10}{\rm Be}$ is shown.
 Panels (d)-(f) show the decay to the excited band of $^{10}{\rm Be}$ is shown.}  
 \label{fig:width2-} 
\end{figure*}

\begin{table*}[h]
 \caption{Partial decay widths (keV) of linear-chain bands for (a) positive-parity and (b) negative-parity linear-chain band 2. The channel radii $a$ are (a) 6.0 fm and (b) 7.0 fm, respectively.}  
\label{tab:lc}
\begin{center}
 \begin{ruledtabular}
  \begin{tabular}{lccccc} 
  \multicolumn{6}{c}{(a) positive-parity} \\
  \hline
   $J^\pi$ & $E_x$ & $\Gamma_\alpha({}^{12}{\rm Be}(0^+_1))$ & $\Gamma_\alpha({}^{12}{\rm Be}(2^+_1))$ & $\Gamma_{^6{\rm He}}({}^{10}{\rm Be}(0^+_1))$ & $\Gamma_{^6{\rm He}}({}^{10}{\rm Be}(2^+_1))$ \\
   \hline
   $0^+_6$ & 16.81 & 335 & 1 & - & - \\
   $2^+_9$ & 17.51 & 300 & 118 & 0 & - \\
   $4^+_{10}$ & 18.99 & 505 & 954 & 33 & - \\
   $6^+_5$ & 21.49 & 535 & 1591 & 78 & 18 \\ \hline
  \multicolumn{6}{c}{(b) negative-parity} \\
  \hline
   $1^-_{11}$ & 22.05 & 198 & 567 & 77 & 63 \\
   $3^-_{14}$ & 23.00 & 196 & 597 & 84 & 115 \\
   $5^-_{15}$ & 24.76 & 181 & 615 & 92 & 173 \\
   $7^-_6$ & 27.35 & 224 & 763 & 100 & 225 \\ 
  \end{tabular}
 \end{ruledtabular}
 \end{center}
\end{table*}

Figure \ref{fig:width1} shows the $\alpha$ reduced widths of several selected low-spin states.
For positive parity, we show the member states of the ground, triangular and linear-chain bands, while for the negative parity, we show only the states which have the reduced widths larger than 0.1 MeV$^{1/2}$.
The decay channels are indicated as $[^{12}{\rm Be}(j^\pi)\otimes l]$ where $j^\pi$ and $l$ denote the angular momentum of the $^{12}{\rm Be}$ ground band and the relative angular momentum between $^{12}{\rm Be}$ and $\alpha$ particles, respectively.
Here, $^{12}$Be is assumed to have two neutrons in $\pi$-orbit and the other two neutrons in $\sigma$-orbit.
The channel radii $a$ are 6.0 fm for (a)-(c) and 5.5(left side), 7.0(right side) fm for (d)-(f), which are chosen to be smoothly connected to the Coulomb wave function.
The detailed values of $\alpha$ and $^{6}{\rm He}$ decay widths for linear-chain states are listed in Tab. \ref{tab:lc}.

In the positive parity, the linear-chain band (the $0^+_6$, $2^+_9$, and $4^+_{10}$ states) has large reduced widths compared to the ground and the triangular bands.
It is also noted that the $\alpha$ reduced widths of other excited states are also much smaller than the linear-chain band.
Hence, in the calculated energy region, the linear chain band has the largest reduced widths.
Another point to be noted is the decay pattern of the linear-chain band.
The reduced widths in the [$^{12}$Be$(2^+_1)\otimes l$] channels are as large as or even larger than those in the [$^{12}$Be$(0^+_1)\otimes l$] channel.
This dominance of the $^{12}{\rm Be}(2^+_1)$ component in the linear-chain band is owe to the strong angular correlation between $\alpha$ clusters which is brought about by their linear alignment.
This property is in contrast to the Hoyle state where $\alpha$ particles are mutually orbiting with $l=0$, and hence, the $^{8}{\rm Be}(0^+_1)$ component dominates \cite{funa15}.
Similar properties of the linear-chain configuration was also discussed in $^{12}{\rm C}$\cite{suzu72} and $^{14}{\rm C}$ \cite{baba16}.
Therefore, if the decay to $^{12}{\rm Be}(2^+_1)$ is confirmed, it will be a strong evidence for the linear-chain formation.

Figure \ref{fig:width2+} shows the $^{6}{\rm He}$ reduced widths of linear-chain states for positive-parity.
We calculated the $^{6}{\rm He}$ reduced widths for both ${}^{6}{\rm He}+{}^{10}{\rm Be}(0^+_1, 2^+_1)$ and ${}^{6}{\rm He}+{}^{10}{\rm Be}(0^+_2, 2^+_3)$ channels.
Here we assumed that the ground band of $^{10}{\rm Be}$($0^+_1$ and $2^+_1$) has the $\pi$-orbit neutrons, while the excited state ($0^+_2$, $2^+_3$, and so on) of $^{10}{\rm Be}$ has the $\sigma$-orbit neutrons.
Fig. \ref{fig:width2+}(a) corresponds to the decay to ${}^{6}{\rm He}+{}^{10}{\rm Be}(0^+_1, 2^+_1)$ and the panel (b) corresponds to the decay to ${}^{6}{\rm He}+{}^{10}{\rm Be}(0^+_2, 2^+_3)$, respectively.
Although the magnitude of the $^{6}{\rm He}$ reduced widths are about a factor of 2 smaller than that of $\alpha$ reduced widths, they are still sufficiently large compared to any other excited states.
It is also noted that the magnitudes of  ${}^{6}{\rm He}+{}^{10}{\rm Be}(0^+_1, 2^+_1)$ and ${}^{6}{\rm He}+{}^{10}{\rm Be}(0^+_2, 2^+_3)$ reduced widths are almost the same order.
This is caused by the unique configuration of linear-chain state in $^{16}{\rm C}$.
The linear-chain configuration in $^{16}{\rm C}$ has the two $\pi$-orbit neutrons and two $\sigma$-orbit neutrons, hence, the linear-chain configuration of $^{16}{\rm C}$ can decay into both $^{10}{\rm Be}(0^+_1, 2^+_1)$ and $^{10}{\rm Be}(0^+_2, 2^+_3)$.
The results shown in Fig. \ref{fig:width2+} is consistent with this explanation.
This decay property should be compared with that of the linear-chains in $^{14}{\rm C}$.
As already discussed in our previous paper \cite{baba16}, the $\pi$-bond linear-chain state of $^{14}{\rm C}$ decays into the $^{10}{\rm Be}(0^+_1, 2^+_1)$ dominantly, while the decay to $^{10}{\rm Be}(0^+_2, 2^+_3)$ is suppressed.
In contrast to the $\pi$-bond linear-chain, the $\sigma$-bond linear-chain state of $^{14}{\rm C}$ decays into the $^{10}{\rm Be}(0^+_2, 2^+_3)$ dominantly and the decay to $^{10}{\rm Be}(0^+_2, 2^+_3)$ is suppressed.

For the negative parity, it can be seen that the linear-chain band 1 and 2 show relatively smaller reduced widths compared to the positive-parity linear-chain band.
The linear-chain configurations of negative parity do not match to the ${}^{4}{\rm He}+{}^{12}{\rm Be}(0^+_1, 2^+_1)$ configuration due to the existence of the valence neutron which occupies the {\it ungerade} $\pi^+_{3/2}$ (see Fig. \ref{fig:density-}).
Therefore, the decay to the ${}^{4}{\rm He}+{}^{12}{\rm Be}(0^+_1, 2^+_1)$ channel is suppressed.
The characteristic decay patterns of the linear-chain configuration can be also seen in negative parity, namely, the reduced widths in the [$^{12}$Be$(2^+_1)\otimes l$] channels are larger than [$^{12}$Be$(0^+_1)\otimes l$] channels.
In addition, the partial decay widths of the linear-chain band 2 listed in Tab. \ref{tab:lc} (b) are very large because of their high excitation energies.
Therefore, if it is observed, the linear-chain formation in the negative-parity can be supported strongly.
However, it is not easy to distinguish the linear-chain band 1 and 2 from $\alpha$ reduced widths because they are almost the same magnitude.

Figure. \ref{fig:width2-} shows the ${}^{6}{\rm He}$ reduced widths of negative-parity linear-chain states for both ${}^{10}{\rm Be}(0^+_1, 2^+_1)$ and ${}^{10}{\rm Be}(0^+_2, 2^+_3)$ channels.
It is interesting that a characteristic difference between the linear-chain band 1 and 2 appears in the ${}^{6}{\rm He}$ reduced widths.
The linear-chain band 2 has the same magnitude of the ${}^{6}{\rm He}$ reduced widths as the $\alpha$ reduced widths.
In addition, the ${}^{6}{\rm He}$ reduced widths of the linear-chain band 2 are even larger than those of positive-parity linear-chain band (see Fig.\ref{fig:width2+}).
On the other hand, the linear-chain band 1 has the smaller ${}^{6}{\rm He}$ reduced widths, especially it hardly decays into $^{6}{\rm He}+^{10}{\rm Be}(0^+_2, 2^+_3)$.
This characteristic difference enables to distinguish the linear-chain bands 1 and 2.

A high-lying excite state at 20.6 MeV was observed by the breakup of $^{6}{\rm He}+{}^{10}{\rm Be}$ \cite{dell16}.
Since the spin-parity was not assigned, several calculated excited states can be the candidate of observed high-lying state, including the linear-chain states naturally.
However, the positive- or negative-parity linear-chain states only show the large $^{6}{\rm He}$ reduced widths near 20.6 MeV.
Although further experimental studies are in need, we suggest the linear-chain state as the candidate of observed high-lying state.

\section{SUMMARY}
We discussed the properties of the linear-chain states of $^{16}$C based on the AMD.
Especially, we focused on their decay mode to identify them experimentally.

In the positive-parity, it is shown that the linear-chain configuration has the valence neutrons occupying molecular-orbits $(3/2^-_\pi)^2(1/2^-_\sigma)^2$.
It generates a rotational band built on $0^+$ state at 16.7 MeV and its moment of inertia is estimated as $\hbar/2\Im=$112 keV.
It was shown that the linear-chain states have the large $\alpha$ and $^6$He reduced widths.
In particular, the large $\alpha$ reduced widths in the $\alpha+{}^{12}{\rm Be}(2^+_1)$ channel is a strong evidence for the linear-chain configuration.
In the case of the $^6$He decay, the magnitudes of the reduced decay widths in both ${}^{10}{\rm Be}(0^+_1, 2^+_1)$ and ${}^{10}{\rm Be}(0^+_2, 2^+_3)$ channels are almost same order.
Compared with $^{14}{\rm C}$, this is caused by the unique configuration of linear-chain state in $^{16}{\rm C}$.

In the negative-parity states, we found two types of linear-chain bands.
The first band, which we call the linear-chain band 1, is composed of the linear-chain configuration with the $(3/2^-_\pi)^2(1/2^-_\sigma)(3/2^+_\pi)$ molecular-orbits.
This band is built on $1^-$ states located around 18.5 MeV.
Because of the mixing with non-cluster states and the mixing of $K=0^-$ and $1^-$ components, the member states are fragmented into several states.
The other band, which we call the linear-chain band 2, is built on $1^-$ states located around 22.1 MeV.
Although this band does not have the clear molecular-orbit configuration, the single rotational $K=0$ band is clearly formed with the large moment of inertia $\hbar/2\Im=$98 keV.
The $\alpha$ reduced widths of these two linear-chains are smaller than those of positive-parity linear-chain band, but are sufficiently large to be distinguished from other non-cluster states.
These two linear-chains cannot be distinguish based on the $\alpha$ reduced widths because they are almost same magnitude.
However, the ${}^{6}{\rm He}$ reduced widths of the linear-chain band 2 are larger than those of linear-chain band 1.
We consider that this characteristic difference enables to distinguish the linear-chain bands 1 and 2.

\acknowledgements
One of the authors (T.B.) acknowledges the support by JSPS KAKENHI Grant No. 16J04889.
The other (M.K.) acknowledges the support by the Grants-in-Aid for Scientific Research on Innovative Areas from MEXT (Grant No. 2404:24105008) and JSPS KAKENHI Grant No. 16K05339.

\end{document}